\documentclass[aps,10pt,twocolumn,showpacs]{revtex4}
\usepackage{amssymb}
\usepackage{graphicx,setspace}
\usepackage{amsmath}

\begin{document}
\title{The influence of an electromagnetic field on the wave-current interaction.}
\author{Germain Rousseaux and Philippe Ma{\"{\i}}ssa}
\affiliation{Universit\'{e} de Nice-Sophia Antipolis,
Laboratoire J.-A. Dieudonn\'{e}, UMR CNRS-UNS 6621,\\
Parc Valrose, 06108 Nice Cedex 02, France, European Union.}


\begin{abstract}

We study the propagation of surface waves on a current in the presence of an electromagnetic field. A  horizontal (vertical) field strengthens (weakens) the counter-current which blocks the waves. We compute the phase space diagrams (blocking velocities versus period of the waves) with and without surface tension. Three new dimensionless numbers are introduced to compare the relative strengths of gravity, surface tension and field effects. This work shows the importance of an electromagnetic field in order to design wave-breakers or in microfluidics applications.

\end{abstract}


\maketitle

The influence of a fluid current on the propagation of water waves was first described by Jean de la Fontaine in the fable The Wolf and the Lamb: {\it "How dare you roil my drink ? Your impudence I shall chastise !" "Let not your majesty," the lamb replies, "Decide in haste or passion! For sure It's difficult to think In what respect or fashion My drinking here could roil your drink, Since on the stream your majesty now faces I'm lower down, full twenty paces"}. Since then, wave-breakers are used to stop gravity waves in marine applications \cite{Taylor}: a curtain of air bubbles blown up from the seabed through a row of perforated nozzles acts as a barrier to the movement of waves over the surface. The rising bubbles generate streams flowing on the surface and blocking the waves. 

Here, following our recent works on the wave-current interaction problem \cite{NJP, Nardin, Rousseaux}, we show that a vertical (horizontal) field can reinforce (diminish) the blocking of water waves by a counter-current. In addition, this study can have applications in microfluidics for the design of electromagnetic actuators where waves are often used to enhance transfer between samples at interfaces moving relatively. 

Surface waves submitted to either an electric or magnetic field behave in a universal manner \cite{Melcherbook1, Melcherbook2}. Indeed, it is known that the influence of a field (be it electric or magnetic, vertical or horizontal) is described by a quadratic correction for the wave-number in the dispersion relation $\omega ^2 = F(k)$ ($\omega /2\pi$ is the frequency of the waves in the rest frame and $k$ the algebraic wave-number, $k^2=k_{x}^2+k_{y}^2$). 
For example, the dispersion relations for the  electrocapillary (magnetocapillary) waves are ($k>0$),
\begin{itemize}
\item  in a vertical electric \cite{Larmor, Bertein, Dean, Surgy,  Melcherbook2}(magnetic \cite{Cowley,Richter}) field:

\begin{equation}
\omega ^2 \simeq \left( gk - \frac{\epsilon E_z^2}{\rho} k^2+\frac{\gamma}{\rho} k^3 \right)
\end{equation}
\begin{equation}
\omega ^2 \simeq \left( gk-\frac{(\frac{\mu}{\mu _0}-1)^2B_z^2}{\rho (1+\frac{\mu }{\mu _0})\mu}k^2+\frac{\gamma}{\rho} k^3 \right),
\end{equation}

\item  in a  horizontal electric \cite{Melcher, Melcherbook1}(magnetic \cite{Rosensweig}) field:

\begin{equation}
\omega ^2 \simeq \left( gk+ \frac{(\epsilon -\epsilon_0)^2E_x^2}{\rho(\epsilon _0+\epsilon )} k_x^2+\frac{\gamma}{\rho} k^3 \right)
\end{equation}
\begin{equation}
\omega ^2 \simeq \left( gk+ \frac{(\mu -\mu _0)^2B_x^2}{\rho(\mu  _0+\mu )\mu ^2} k_x^2+\frac{\gamma}{\rho} k^3 \right),
\end{equation}
\end{itemize}
where $g$ denotes the gravitational acceleration of the Earth at the water surface, $\rho$ the fluid density, $\gamma$ its surface tension, $\mu$ its permeability, $\epsilon$ its permittivity, $E$ ($B$) an external electric (magnetic) field. All the preceding dispersion relations are valid if the fluid depth $h$ is such that $|k|h \gg1$ (thick case). This hypothesis will be used in the rest of the paper.

Besides, the presence of a uniform current induces a Doppler shift of the pulsation $\omega$  \cite{Dingemans, FS, Igor, Peregrine},
\begin{equation}
(\omega -{\bf U.k})^2 = F(k),
\end{equation}
where $U$ is the constant velocity of the background flow. We will assume that $k=k_x$ and $U=U_x$ for simplicity. The dispersion relation features the following symmetry: $U\to -U$ and $k \to -k$; hence, negative energy waves ($k<0$ and $\omega -{\bf U.k}<0$) are also described using this symmetry \cite{FS}. For the quadratic term, the symmetry is hidden by the deep water approximation ($|k|h \gg1$) where a $coth(kh)\simeq -1$ term introduces a negative sign for negative wavenumbers \cite{Rosensweig}. 

Hence, the universal dispersion relation for water waves propagating on a counter-current in presence of an horizontal ($+v_f^2$) or  vertical ($-v_f^2$) field  writes
\begin{equation} \label{dispersion}
(\omega -Uk)^2 \simeq \lbrack s \rbrack \left( gk + \lbrack s \rbrack (\pm v_{f} ^2) k^2+\frac{\gamma}{\rho} k^3 \right),
\end{equation}
where $\lbrack s \rbrack= \pm 1$ is the sign of $k$. $v_{f}$ can be seen as the effective velocity resulting from the Maxwell electromagnetic tensions $\sigma _{Maxwell} \approx \epsilon E^2$ or $\frac{B^2}{\mu} = \rho  v_{f}^2$ \cite{Melcherbook2}.

First, we study the influence of a field on the propagation of gravity waves on a counter-current ($U<0$). Then, we include the effect of capillarity. The examples shown on the figures correspond to the case of an electric field $E$ (horizontal and/or vertical).

Following our dynamical system approach \cite{Nardin, Rousseaux}, wave blocking can be seen as a saddle-node bifurcation for the wave-number. Indeed, the usual condition  \cite{Dingemans, FS, Igor, Peregrine} for blocking --group velocity vanishes, $\partial \omega /\partial k=0$-- is equivalent to the appearance of a double root in the dispersion relation $P(k)(k-k_2)^2=0$ where $P(k)$ is a polynomial.

First, we neglect surface tension and from (\ref{dispersion}) the dispersion relation leads to a quadratic polynomial in $k$ ($U_*$ is a blocking velocity)
\begin{equation} \label{keqn1a}
k^2-\frac{(\lbrack s \rbrack g+2\omega U_*)}{(U_*^{2} -(\pm v_{f} ^2))} k+\frac{\omega ^2}{(U_*^{2}-(\pm v_{f}^2))}=0,
\end{equation}
with $|U_*|\neq v_{f}$ in the horizontal field case.\\
Close to the bifurcation, we have:
\begin{equation} \label{keqn2a}
(k-k_2)^2=(k^2-2k_2k+k_2 ^2)=0.
\end{equation}
From (\ref{keqn1a}) and (\ref{keqn2a}), we get two expressions for a blocking wave-number with the constraint $|U_*| > v_f$ in the horizontal field case:
\begin{equation}
k_2 = \frac{(\lbrack s \rbrack g+2\omega U_*)}{2(U_*^{2}-(\pm v_{f} ^2))}=\lbrack s \rbrack \frac{\omega}{\sqrt{(U_*^{2} -(\pm v_{f}^2))}}
\end{equation}
so that, with $\omega = 2\pi /T$, a blocking velocity writes:
\begin{equation}
U_*=-\lbrack s \rbrack (\frac{gT}{8\pi}+\frac{2\pi(\pm v_{f} ^2)}{gT})
\end{equation}
and the associated blocking wave-number becomes:
\begin{equation} 
k_2=\lbrack s \rbrack \frac{2\pi}{T\sqrt{\frac{4\pi ^2 v_{f} ^4}{g^2T^2}-(\pm \frac{v_{f} ^2}{2})+\frac{g^2T^2}{64 \pi ^2}}}
\end{equation}

\begin{figure}[!htbp]
\includegraphics[width=8cm]{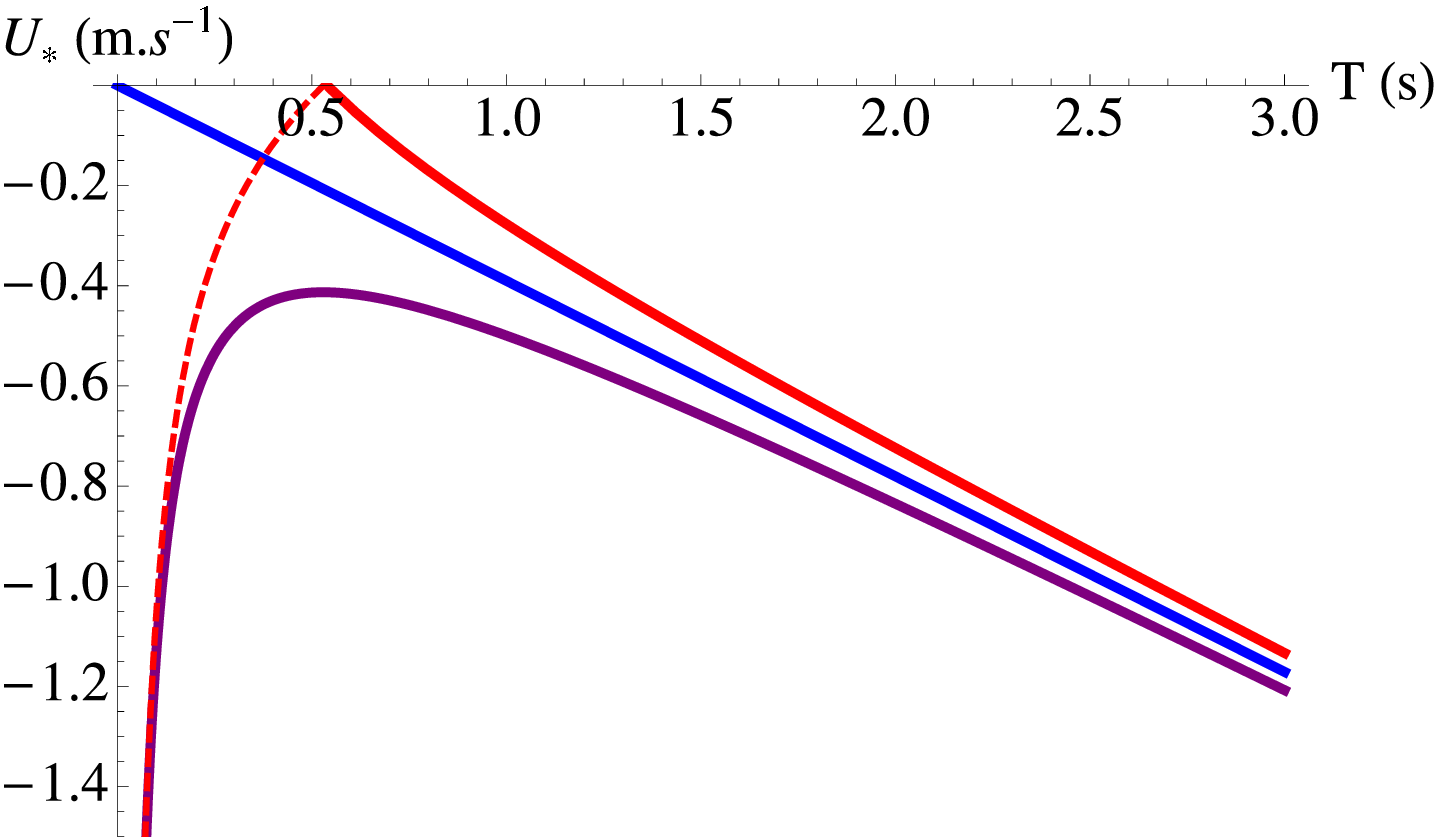}
\caption{Analytical phase space : the blocking velocities $U_*$ versus the period $T$ of incoming waves. $\gamma =0$ $N.m^{-1}$, $\rho = 1000$ $kg.m^{-3}$, $E=5.10^5$ V/m, $\epsilon_r=80$ (water) $\Rightarrow$ $v_{f} \simeq 0.42$ $ms^{-1}$. Blue line : no field (left side : blocking). Red lines : vertical fields (left side : blocking). Dotted red line : threshold for the appearance of negative energy waves. Purple line : horizontal fields (bottom side : blocking).}
\label{phase1}
\end{figure}

We display on Figure \ref{phase1} the phase space of the control parameters $U_*$ and $T$. If $v_{f} =0$, we recover (\lbrack s\rbrack =+1) $U_*=U_g=-\frac{gT}{8\pi}$ and $k_2=\frac{16\pi ^2}{gT^2}$ \cite{Nardin}. When $v_{f} \neq 0$ this blocking line is modified and, only for the vertical field, a new one appears for $k_2 < 0$ (\lbrack s\rbrack =-1). So that, with a vertical field, all the waves with a period inferior to $T_f= \frac{4\pi v_f }{g}$ are blocked even with no counter-flow and there is a threshold for the appearence of negative energy waves for this same range of periods. With an horizontal field, the strengh of the counter-flow to reach wave blocking is strongly increased at low periods and even, blocking will not at all occur provided that $\left| U_*\right| <\left| v_f \right|$.\\

Now, the surface tension modifies the dispersion relation and from (\ref{dispersion}), it becomes a cubic polynomial in $k$:
\begin{equation} \label{keqn1b}
k^3 -\lbrack s \rbrack \frac{\rho (U_*^2 -(\pm v_{f} ^2))}{\gamma}k^2+\frac{\rho}{\gamma}(g+\lbrack s \rbrack 2\omega U_*)k- \lbrack s \rbrack \frac{\rho \omega ^2}{\gamma}=0.
\end{equation}
The condition for wave-blocking becomes:
\begin{equation} \label{keqn2b}
\begin{split}
(k-k_1)(k-k_2)^2 = k^3 -(2k_2+k_1)k^2+\\
(k_2^2+2k_1k_2)k-k_1k_2^2=0.
\end{split}
\end{equation}
From (\ref{keqn1b}) and (\ref{keqn2b}), we end up with a quadratic polynomial for the blocking wave-numbers $k_2$,
\begin{equation} \label{k2eqn}
3k_2^2- \lbrack s \rbrack 2\frac{\rho (U_*^2 -(\pm v_{f} ^2 ))}{\gamma}k_2+\frac{\rho}{\gamma}(g+ \lbrack s \rbrack 2\omega U_*)=0,
\end{equation}
whose determinant is
\begin{equation} \label{deltaeqn}
\Delta =4\frac{\rho ^2 (U_*^2-( \pm v_{f} ^2))^2}{\gamma ^2}-12\frac{\rho}{\gamma}(g+ \lbrack s \rbrack
2\omega U_*),
\end{equation}
and $k_2$ and $k_1$ write,
\begin{equation}  \label{k2}
k_2^{(a, b)}= \lbrack s \rbrack \frac{\rho (U_*^2 -(\pm v_{f} ^2))}{3\gamma} \left(1 \pm\sqrt{1-\frac{}{} \frac{3\gamma(g+\lbrack s \rbrack 2\omega U_*)}{\rho(U_*^2-(\pm v_{f} ^2))^2}} \right)
\end{equation}
\begin{equation}  \label{k1}
k_1^{(a, b)}=\lbrack s \rbrack \frac{\rho (U_*^2 -(\pm v_{f} ^2))}{3\gamma} \left(1 \mp 2\sqrt{1-\frac{}{} \frac{3\gamma(g+\lbrack s \rbrack 2\omega U_*)}{\rho(U_*^2-(\pm v_{f} ^2))^2}} \right)
\end{equation}
with the constraint,
\begin{equation}  \label{constreqn}
k_1k_2^2 = \lbrack s \rbrack \frac{\rho \omega ^2}{\gamma}.
\end{equation}
After tedious algebra, (\ref{constreqn}) leads to the following quintic in $U_*$ with coefficients depending on the period and the field (in addition to the fluid characteristics and the universal constants):
\begin{widetext}
\begin{eqnarray}
	\nonumber
	0 & = & 12\rho ^2 g \omega U_* ^5  +  \lbrack s \rbrack \left(3\rho ^2 g^2 + 12 \rho ^2(\pm v_{f} ^2)\omega ^2\right)U_* ^4
	- \left(24\rho ^2 g (\pm v_{f} ^2) \omega-12\gamma \rho \omega ^3\right)U_* ^3\\
	\nonumber
	  & - & \lbrack s \rbrack \left(24\rho ^2 v_{f} ^4 \omega ^2+6\rho ^2 g^2(\pm  v_{f} ^2)+ 90\gamma \rho g \omega ^2\right)U_* ^2
	  + \left(12 \rho ^2 g v_{f} ^4 \omega - 108 \gamma \rho (\pm v_{f} ^2) \omega ^3 -72\gamma \rho  g^2 \omega \right)U_*\\
	   & + & \lbrack s \rbrack \left(12\rho ^2 \omega ^2 (\pm v_{f} ^6) +3\rho ^2 g^2 v_{f} ^4 - 54 \gamma \rho g (\pm v_{f} ^2) \omega ^2 -81\gamma ^2 \omega ^4 -12\gamma \rho g^3\right).
	\label{quintic}
\end{eqnarray}
\end{widetext}

\begin{figure}[!htbp]
\includegraphics[width=9cm]{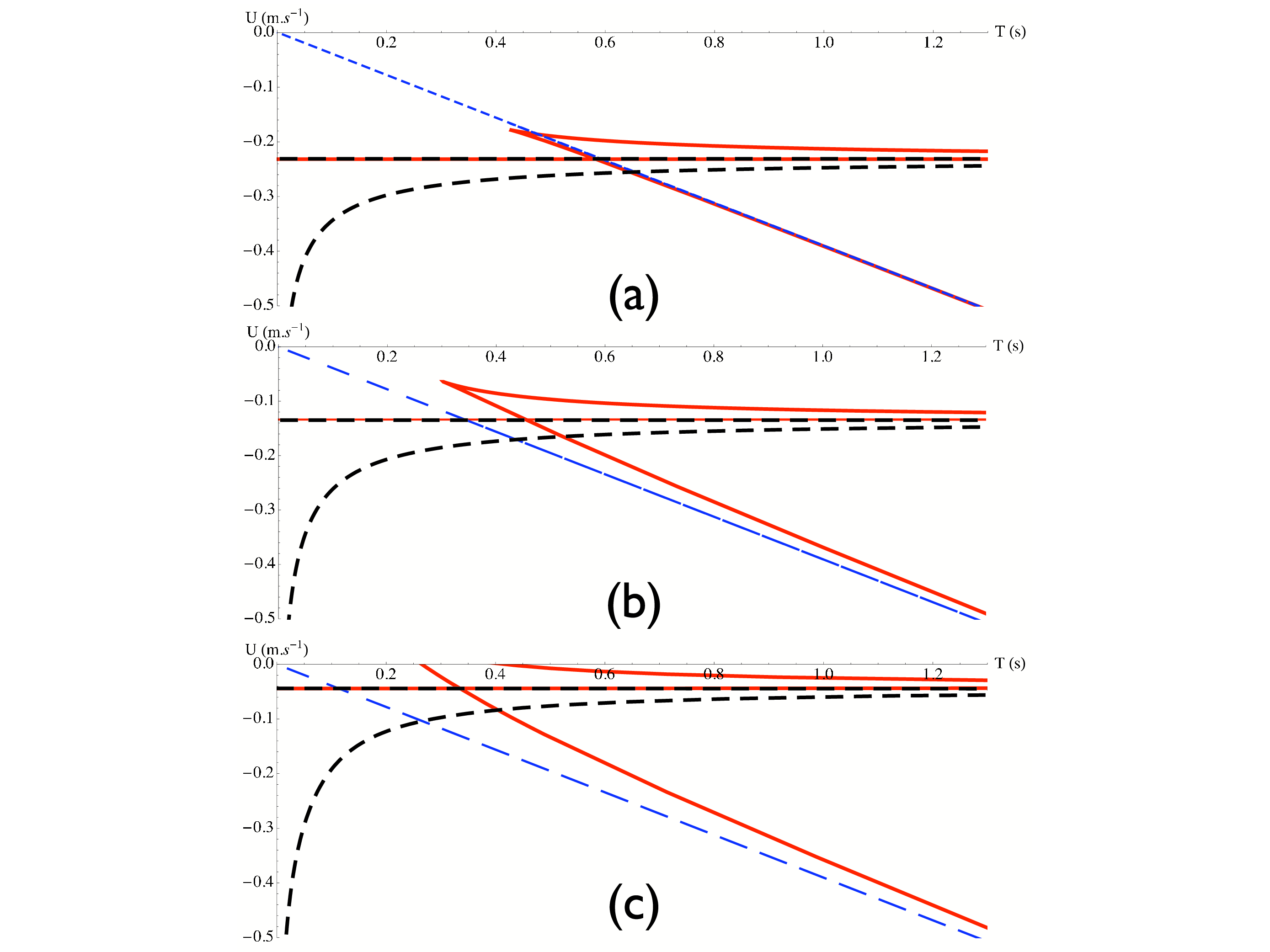}
\caption{Numerical phase space (vertical fields): the blocking velocities $U_*$ versus the period $T$ of incoming waves. $\gamma =0.073$ $N.m^{-1}$, $\rho = 1000$ $kg.m^{-3}$, $\epsilon _r=80$ (water). (a) $E_z=0$ $V/m$, (b) $E_z=2.24$ $10^{5}$ $V/m$, (c) $E_z=2.7$ $10^{5}$ $V/m$. Dotted blue line: $U_*=U_g$ no field and no surface tension (left : blocking). Red lines : upper = threshold for the appearance of positive energy waves = blue-shifted waves blocking boundary ; lower = gravity waves blocking boundary. Dotted black line : threshold for the appearance of negative energy waves (bottom). Dotted red and black line: $U_*=U_a$.}
\label{phase2}
\end{figure}

The quintic can be solved numerically as in \cite{Rousseaux}, but here we have used an implicit method.
We computed numerically from the dispersion relation the associated phase space to the quintic in a parametric plot --$U_* $ versus $T$-- using $k$ as the hidden variable (Fig.  \ref{phase2}). Indeed, the dispersion relation (\ref{dispersion}) is written as $(\omega-U_*k) =\pm f(k)$. Hence, we plot the parametric equation $(T(k), U_*(k))=\left(\pm (\frac{2\pi}{f(k)-f'(k)k}), \mp f'(k) \right)$ for a range of given $k$.

It is interesting to look at particular cases and asymptotic behaviours starting from the quintic to recover some of the numerical results.

\begin{itemize}
\item{$v_{f} =0$:}
\begin{equation}
\begin{split} 
12\rho ^2 g \omega (U_* ^5+\lbrack s \rbrack \frac{1}{4}\frac{g}{\omega}U_* ^4+\frac{\gamma \omega ^2 }{\rho g}U_* ^3-\lbrack s \rbrack \frac{15}{2}\frac{\gamma \omega }{\rho }U_* ^2-6\frac{g\gamma  }{\rho }U_*\\
-\lbrack s \rbrack (\frac{\gamma g^2 }{\rho \omega}+\frac{27}{4}\frac{\gamma ^2 \omega ^3 }{\rho ^2 g})) =0
\end{split} 
\end{equation}
In the absence of any field, we recover the quintic polynomial  and the phase space (Fig.\ref{phase2}(a)) derived very recently in \cite{Rousseaux}.
\item{$\omega \to 0$:}
For large period, we find a quartic polynomial in $U_*$.
\begin{equation} 
\left(U_* ^4-2 (\pm v_{f} ^2) U_* ^2 +v_{f} ^4 -4\frac{\gamma g}{\rho}\right)=0,
\end{equation}
whose  solution is the asymptotic velocity $U_a$:
\begin{equation} \label{Ua}
\lim\limits_{T\to \infty} U_* = U_a =U_\gamma \sqrt{1 +\frac{(\pm v_{f} ^2)}{U_\gamma ^2}}
\end{equation}
where $U_\gamma = -\sqrt{2}\left(\frac{\gamma g}{\rho}\right)^{1/4}$ is  the asymptotic velocity when $v_f$ = 0 (see (\cite{Rousseaux}).  From (\ref{k2}), we get the associated wave numbers, which is independent of the field,  $k_a = \lbrack s \rbrack (\frac{\rho g}{\gamma})^{1/2} = k_\gamma$ (\cite{Rousseaux}). These asymptotic behaviors can be seen in (Fig.\ref{phase2}).

\item{$U_* \to \infty$:}
The quintic becomes:
\begin{equation}
12\rho ^2 g \omega U_* ^5 + \lbrack s \rbrack (3\rho ^2 g^2 + 12 \rho ^2(\pm v_{f} ^2) \omega ^2)U_* ^4 \simeq 0
\end{equation}
Hence, the other asymptotic limit is:
\begin{equation} \label{U*inf}
U_* \simeq - \lbrack s \rbrack  \left(\frac{g}{4\omega}+ \frac{\omega (\pm v_{f} ^2)}{g}\right)= \lbrack s \rbrack U_g \left(1+\frac{1}{4}\frac{(\pm v_{f} ^2)}{U_{g}^2}\right)
\end{equation}
which describes how the lower red line (blocking threshold in the presence of a field) deviates from the dotted blue line (pure gravity regime) in (Fig.\ref{phase2}).

\end{itemize}
We introduce three new dimensionless numbers $R_{gf}$, $R_{\gamma f}$ and $R_{g \gamma}$ :
\begin{equation}
R_{gf} =\frac{U_g}{v_{f}} \quad R_{\gamma f} =\frac{U_\gamma}{v_{f}} \quad R_{g \gamma}=\frac{R_{gf}}{R_{\gamma f} }=\frac{U_g}{U_\gamma}
\end{equation}

If $R_{gf} \gg1$, then from (\ref{U*inf}), $U_* \to U_g$ \cite{Nardin}. $R_{gf}$ compares the effect of gravity with the field effect. If $R_{\gamma f} \gg1$, then from (\ref{Ua}), $U_{a} \to U_\gamma $ \cite{Rousseaux}. $R_{\gamma f}$ compares the effect of surface tension with the field effect. $R_{g \gamma}$ compares the effect of gravity with surface tension. For example, waves will be of the capillary type if $R_{g \gamma} \ll1$.

The critical point, the cusp \cite{Rousseaux}, is reached when the determinant $\Delta$ of the quadratic polynomial (\ref{k2eqn}) equals zero, so that, $k_2^{(a,b)}=k_1^{(a,b)}=k_c$. From (\ref{k2}) (or (\ref{k1})) and from the constraint (\ref{constreqn}), we get respectively,
\begin{equation}
k_c = \lbrack s \rbrack \frac{1}{3} \frac{\rho (U_c ^2 -(\pm v_{f} ^2))}{\gamma}=  \lbrack s \rbrack \left(\frac{\rho \omega _c ^2}{\gamma}\right)^{1/3},
\end{equation}
so that the velocity $U_c$ at the cusp writes,
\begin{equation}
U_c = -\sqrt{3\left(\frac{\gamma \omega _c }{\rho }\right)^{2/3}+(\pm v_{f} ^2)},
\end{equation}
and the determinant  $\Delta$ (\ref{deltaeqn}) 
at the cusp leads to an equation for $\omega _c$:
\begin{equation}
\begin{split}
3\left(\frac{\gamma  \omega _c}{\rho }\right)^{8/3}+6\left(\frac{\gamma g}{\rho }\right)\left(\frac{\gamma  \omega _c}{\rho }\right)^{4/3}+4(\pm v_{f} ^{2})\left(\frac{\gamma  \omega _c}{\rho }\right)^{2}\\
-\left(\frac{\gamma  g }{\rho }\right)^{2}=0.
\end{split}
\end{equation}
Using $X=\left(\frac{\gamma \omega _c }{\rho }\right)^{2/3}$, we get a quartic in $X$,\\
\begin{equation}
X^{4} +\frac{4}{3}(\pm v_{f}^{2})X^{3}+2\left(\frac{\gamma g }{\rho}\right)X^{2}-\frac{1}{3}\left(\frac{\gamma g}{\rho}\right)^{2}=0,
\end{equation}
which can be solved numerically. An analytical but lengthy expression for $X$ as a function of $\left(\frac{\gamma g }{\rho}\right)$ and ($\pm v_f ^2$) has also been found but is out of the scope of this letter. We shall come back to this point in a forthcoming paper.

Finally, $\omega_c$, $k_c$ and $U_c$ respectively write as a function of $X$:
\begin{equation}
\begin{split}
\omega_c = (\frac{\rho}{\gamma})X^{3/2}, \quad k_c = \lbrack s \rbrack (\frac{\rho}{\gamma})X,\\
U_c = -\sqrt{3X+(\pm v_{f}^2)}
\end{split}
\end{equation}

In this work, we have shown the influence of a field (electric or magnetic) on the interaction between surface waves and a current. It should be interesting to test experimentally the "damping" effect of a vertical electric field on the propagation of waves in the design of wave-breakers. For microfluidics applications, a generalization of our results  encoding different fluid densities would include another dimensionless quantity, the Atwood number $A_t =\frac{\rho_2 -\rho_1}{\rho_1+\rho_2}$.

\section*{Acknowledgements}
The authors would like to thank Harunori Yoshikawa for fruitful discussions.

\end{document}